\documentclass[%
 reprint,
 amsmath,amssymb,
 aps,
 twocolumn,
]{revtex4}

\usepackage{graphicx}% Include figure files
\usepackage{dcolumn}% Align table columns on decimal point
\usepackage{bm}% bold math
\usepackage{graphicx}
\usepackage{mathtools}
\usepackage{braket}
\usepackage{stackengine}

\newcommand\figref{\figurename~\ref}
\newcommand{\ketr}[1]{\left| #1 \right\rangle}
\newcommand{\brar}[1]{\left\langle #1 \right|}
\begin{document}

\preprint{APS/123-QED}

\title{Multi-Photon Synthetic Lattices in Multi-Port Waveguide Arrays: Synthetic Atoms and Fock Graphs}

\author{Konrad Tschernig}
\affiliation{Max-Born-Institut\\
Max-Born-Stra{\ss}e 2A\\
12489 Berlin, Germany }
\email{konrad.tschernig@mbi-berlin.de}

\author{Roberto de J. Le\'on-Montiel}%
 \affiliation{Instituto de Ciencias Nucleares\\
	Universidad Nacional Aut\'onoma de M\'exico\\
	Apartado Postal 70-543, 04510 Cd. Mx., M\'exico
}%

\author{Armando Perez-Leija}
\affiliation{Max-Born-Institut\\
Max-Born-Stra{\ss}e 2A\\
12489 Berlin, Germany 
}%

\author{Kurt~Busch}
\affiliation{%
Humboldt-Universit\"at zu Berlin\\
Institut f\"{u}r Physik, AG Theoretische Optik \& Photonik\\
Newtonstra{\ss}e 15, 12489 Berlin, Germany\\
}%

%\author{Konrad Tschernig\thanks{Corresponding author: konrad.tschernig@mbi-berlin.de} \\
%	Max-Born-Institut\\
%	Max-Born-Stra{\ss}e 2A\\
%	12489 Berlin, Germany \\
%	%% examples of more authors
%	\And
%	Roberto de J. Le\'on-Montiel \\
%	Instituto de Ciencias Nucleares\\
%	Universidad Nacional Aut\'onoma de M\'exico\\
%	Apartado Postal 70-543, 04510 Cd. Mx., M\'exico
%	\And
%	 Armando Perez-Leija \\
%	 Max-Born-Institut\\
%	Max-Born-Stra{\ss}e 2A\\
%	12489 Berlin, Germany \\
%	 \And
%	 Kurt~Busch \\
%	 Humboldt-Universit\"at zu Berlin\\
%	 Institut f\"{u}r Physik, AG Theoretische Optik \& Photonik\\
%	 Newtonstra{\ss}e 15, 12489 Berlin, Germany
%	%% \And
%	%% Coauthor \\
%	%% Affiliation \\
%	%% Address \\
%	%% \texttt{email} \\
%}

\date{\today}% It is always \today, today,

\begin{abstract}
Activating transitions between internal states of physical systems has emerged as an appealing 
approach to create lattices and complex networks. In such a scheme, the internal states or modes 
of a physical system are regarded as lattice sites or network nodes in an abstract space whose
dimensionality may exceed the systems' apparent (geometric) dimensionality. This introduces  
the notion of synthetic dimensions, thus providing entirely novel pathways for fundamental 
research and applications.  
Here, we analytically show that the propagation of multi-photon states through multi-port 
waveguide arrays gives rise to synthetic dimensions where a single waveguide system generates 
a multitude of synthetic lattices. Since these synthetic lattices exist in photon-number space, 
we introduce the concept of pseudo-energy and demonstrate its utility for studying multi-photon 
interference processes. Specifically, the spectrum of the associated pseudo-energy operator 
generates a unique ordering of the relevant states. Together with generalized pseudo-energy 
ladder operators, this allows for representing the dynamics of multi-photon states by way of 
pseudo-energy term diagrams that are associated with a \emph{synthetic atom}. 
As a result, the pseudo-energy representation leads to concise analytical expressions for 
the eigensystem of $N$ photons propagating through $M$ nearest-neighbor coupled waveguides. 
In the regime where $N>2$ and $M>2$, non-local coupling in Fock space gives rise to hitherto
unknown all-optical dark states which display intriguing non-trivial dynamics.
\end{abstract}

\maketitle

%\tableofcontents

\section{Introduction}
\begin{figure}
   \centering
   \includegraphics[scale=.13]{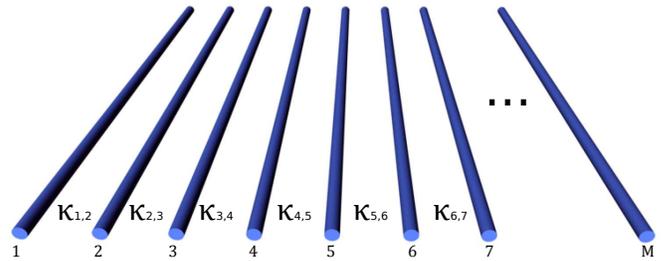}
      \caption{One-dimensional array of $M$ identical nearest-neighbour evanescently 
               coupled waveguides with coupling coefficients $\kappa_{m,m+1}$. }
   \label{Fig:SArray}
\end{figure}

The concept of synthetic dimensions has recently opened the door to novel perspectives for 
expanding the dimensionality of well-understood physical systems 
\cite{Juki2013,Yuan2016,Bilitewski2016,Ozawa2016,Fan2017}. 
One strategy to explore synthetic dimensions consists in driving the associated dynamical 
systems in order to activate the coupling between different internal modes which under normal 
conditions remain uncoupled 
\cite{Yuan2018}. 
By doing so, the resulting coupled modes exhibit lattice-like structures that exist in an 
abstract space which is nonetheless physical. 
The importance of synthetic lattices lies on the fact that they allow us to explore a variety 
of effects that are not available in spatial or temporal domains.\\
To illustrate the basic idea of activating synthetic dimensions, and to set the stage for
the present work, we begin by elucidating how a one-dimensional quantum harmonic oscillator
generates a lattice in Fock space. 
The oscillator's Hamiltonian is given as $\hat{H}=\omega\left(\hat{a}^{\dagger}\hat{a}+\frac{1}{2}\right)$, 
and its dynamics is governed by the Schr\"odinger equation $i\partial_t\ket{\Psi(t)} = \hat{H}\ket{\Psi(t)}$. 
Here, $\omega$ is the angular frequency of the oscillator, and $\hat{a}$ and $\hat{a}^{\dagger}$
denote, respectively, the annihilation and creation operators
\cite{Louisell}. 
Notice, we have set the reduced Planck constant and the oscillator mass to unity, 
i.e., $\hbar =1$ and $m_o=1$. When the oscillator is initially prepared in the eigenstate 
$\ket{\Psi(0)}=\ket{n}$, then it will remain in this state, only acquiring a time-dependent
phase factor during evolution, i.e., $\ket{\Psi(t)}=\text{e}^{-i\left(n+\frac{1}{2}\right)\omega t}\ket{n}$.
No transitions to other eigenstates occur. 
However, by subjecting the oscillator to a time-dependent displacement, 
$\hat{x}(t)=f(t)\left(\hat{a}^{\dagger}+\hat{a}\right)$, the Hamiltonian acquires 
the form  $\hat{H}(t)=\omega\left(\hat{a}^{\dagger}\hat{a}+\frac{1}{2}\right)+f(t)\left(\hat{a}^{\dagger}+\hat{a}\right)$. 
Substituting the general state vector $\ket{\Psi(t)}=\sum_{m=0}^{\infty}c_{m}(t)\ket{m}$ - 
where $c_{m}(t)=\bra{m}\hat{U}(t)\ket{\Psi(0)}$ are the transition amplitudes from the initial 
state $\ket{\Psi(0)}$ to the final state $\ket{m}$ and $\hat{U}(t)$ is the time evolution operator - into the Schr\"odinger equation, we find 
that the amplitudes $c_{m}(t)$ obey the semi-infinite set of coupled differential equations 
\begin{align}
   i\frac{dc_{0}}{dt} &=  f(t)c_{1}(t),\\
   i\frac{dc_{m}}{dt} &= \omega m c_{m}(t) + f(t)\left(\sqrt{m}c_{m-1}(t)+\sqrt{m+1}c_{m+1}(t)\right).
\label{eq:Coupled}
\end{align}
These equations clearly illustrate that the time dependent displacement $\hat{x}(t)$ 
activates transitions between the amplitudes $c_{m-1}(t)$, $c_{m}(t)$, and $c_{m+1}(t)$. 
This implies that in Fock space the oscillator generates a lattice, where it can 
"hop" from eigenstate $\ket{m}$ to the adjacent eigenstates $\ket{m-1}$ and $\ket{m+1}$ 
with \emph{hopping rates} $f(t)\sqrt{m}$ and $f(t)\sqrt{m+1}$, respectively 
\cite{PhysRevA.95.023607,Apleija2010,Keil2011,Apleija2012,Keil2012,Nezhad2013,Wang2016}.\\
In general, applying dynamic modulations to the potentials associated with physical 
systems induces coupling among the supported eigenstates. Using this technique, a 
photonic topological insulator in synthetic dimensions has been recently implemented 
via modulated waveguide lattices \cite{Eran,Ozawa2019}. 
Synthetic dimensions have also been explored in harmonic traps \cite{Price2017}, 
optical lattices \cite{Wang2015}, cavities \cite{Wang2016} and even in room-temperature Rydberg atoms \cite{Cai2019}.\\ 
Within the realm of optics and photonics, synthetic dimensions can be created by 
exploiting the spatial, temporal, polarization, and frequency degrees of freedom 
of light \cite{Yuan2018}. 
For instance, large-scale parity-time symmetric lattices have been implemented in 
the temporal domain using optical fiber loops endowed with gain and loss \cite{Regensburger2,Regensburger1}
and a driven-dissipative analogon of the four-dimensional quantum Hall effect has 
been observed in a spatially 3D resonator lattice \cite{Ozawa2016}.\\
In this work, we show that high-dimensional lattices emerge in photon-number space 
when a photonic lattice of $M$ ports \cite{Longhi2009,Szameit2010} 
is excited by $N$ indistinguishable photons, see Fig.~(\ref{Fig:SArray}). 
More precisely, the Fock-representation of $N$-photon states in systems composed 
of $M$ evanescently coupled single-mode waveguides yields to a new layer of 
abstraction where the associated states can be visualized as the energy levels 
of a \emph{synthetic atom} that features a number of allowed and disallowed
transitions between its energy levels.\\
In photonic waveguide lattices, where all the waveguides are coupled to each other, the quantum optical Hamiltonian in paraxial approximation is given as 
$\hat{H} = \sum_{j=1}^M \beta_j \hat{a}_j^\dagger\hat{a}_j  
           + \sum_{\substack{i\neq j}}^M \kappa_{ij} \hat{a}^\dagger_i \hat{a}_j,
$\cite{Chak2007}, 
where $\hat{a}^\dagger_j$ and $\hat{a}_j$, respectively, are bosonic creation 
and annihilation operators for photons in the $j$-th waveguide. Further, 
$\beta_j$ denotes the propagation constant of the $j$-th waveguide and 
$\kappa_{ij}$ is the coupling coefficient between the $i$-th and $j$-th 
waveguide.\\ 
For simplicity we restrict our subsequent analysis to the simplest scenario of 
(in real space) essentially one-dimensional waveguide arrays with nearest-neighbor 
couplings
\begin{equation}
   \hat{H} = \sum_{j=1}^M \left[\beta_j \hat{a}_j^\dagger\hat{a}_j + \kappa_{j,j-1} \hat{a}^\dagger_{j-1} \hat{a}_j + \kappa_{j,j+1} \hat{a}^{\dagger}_{j+1} \hat{a}_j \right].
\label{eq:Ham}
\end{equation}
Under these premises, the propagation of a single-photon along the waveguide can
be described using the Heisenberg equations of motion for the bosonic creation 
operators 
\cite{Lai1991,Bromberg2009}
\begin{align}
   i\frac{d\hat{a}^{\dagger}_{m}}{dz} = \beta_{m} \hat{a}^{\dagger}_{m} + \kappa_{m,m-1}\hat{a}^{\dagger}_{m-1}+ \kappa_{m,m+1}\hat{a}^{\dagger}_{m+1},
\label{eq:Heis}
\end{align}
where $m = 1, \ldots, M$.
Accordingly, the single-photon response is computed through the input-output transformation 
$\hat{a}^{\dagger}_{m}(0) = \sum_{n=1}^{M}U_{m,n}(z)\hat{a}^{\dagger}_{n}(z)$, 
where 
$U_{m,n}(z)$ denotes the $(m,n)$ matrix element of the evolution operator 
$\hat{U}(z)=\exp\left(-iz\hat{H}\right)$ \cite{Markus2016}. 
Using this formalism, it is straightforward to show that an initial $N$-photon state 
$\ket{n_{1},n_{2},...,n_{M}}$, with $N=\sum_{m=1}^{M}n_{m}$, will transform into 
the output state
\begin{align}
    & \ket{\Psi(0)} = \frac{\left(\hat{a}^{\dagger}_{1}(0)\right)^{n_1} \ldots \left(\hat{a}^{\dagger}_{M}(0)\right)^{n_M}}{\sqrt{n_1! \ldots n_M!}}\ket{0}\overset{z}{\longrightarrow}\\
   & \frac{\left(\sum_{n=1}^{M}U_{1,n}(z)\hat{a}^{\dagger}_{n}(z)\right)^{n_{1}}\ldots  \left(\sum_{n=1}^{M}U_{M,n}(z)\hat{a}^{\dagger}_{n}(z)\right)^{n_{M}}}{\sqrt{n_1!\ldots n_M!}}\ket{0}.\nonumber
\label{eq:ResN}
\end{align}
In the context of waveguide lattices, the input-output formalism is by far the most common approach used to compute the output states \cite{MarkusReview}. 
Nonetheless, as we will demonstrate in the remainder of the manuscript, the input-output scheme fails to expose the intrinsic coupling interactions between 
the emerging states.\\
In what follows, we use the equivalent Schr\"odinger-picture formalism to unveil 
the high-dimensional lattice structures arising from the propagation of multiple 
photons through multi-port waveguide systems. To do so, we first notice that $N$ 
indistinguishable photons exciting $M$ coupled waveguides, give rise to a total 
of $N_F = (N+M-1)!/N!(M-1)!$ states which are given by all permutations of the 
integer partitions of $N$ among the $M$ sites.\\
For the trivial case of $N=1$ photon, we simply obtain a set of $M$ states
\begin{equation} \label{eq:1-M}
   \ketr{1_{m}} = \ket{0,\ldots,\underbrace{1}_{\mathclap{m\text{'th waveguide}}},\ldots,0},
\end{equation}
with $m=1,\ldots,M$. By computing the matrix elements of the Hamiltonian given 
in \eqref{eq:Ham} for $N=1$, 
$H_{n,m} = \brar{1_{n}}\hat{H}\ket{1_{m}} 
        = \beta_{n}\delta_{m,n}+\kappa_{n,m-1}\delta_{n,m-1}+\kappa_{n,m+1}\delta_{n,m+1}$, 
one can readily see that the single-photon states are coupled to each other as 
displayed by the equations
\begin{align}
   i\frac{d }{dz}\ket{1_{m}} = \beta_{m} \ket{1_{m}} + \kappa_{m,m-1}\ket{1_{m-1}}+\kappa_{m,m+1}\ket{1_{m+1}},
\label{eq:SPhoton}
\end{align}
in agreement with \eqref{eq:Heis}.\\
We now consider the more interesting scenario of $N$ photons propagating through a 
waveguide beam splitter, $M=2$, with propagation constants $\beta_{1}$ and $\beta_{2}$
and symmetric coupling, i.e., $\kappa_{1,2} = \kappa_{2,1} \equiv \kappa$. 
In this case, there exists a total of $(N+1)$ states, namely $\left(\ket{0,N},\ket{1,N-1},...,\ket{N-1,1},\ket{N,0}\right)$, 
and the Hamiltonian given in \eqref{eq:Ham} acquires the form 
\begin{align}
    \hat{H} =   \beta_{1} \hat{a}_1^\dagger\hat{a}_1 + \beta_{2} \hat{a}_2^\dagger\hat{a}_2 + \kappa \hat{a}^\dagger_{1} \hat{a}_2 + \kappa \hat{a}_1\hat{a}^{\dagger}_{2}. 
\end{align}
\begin{figure}[t!]
\centering
   \includegraphics[scale=.29]{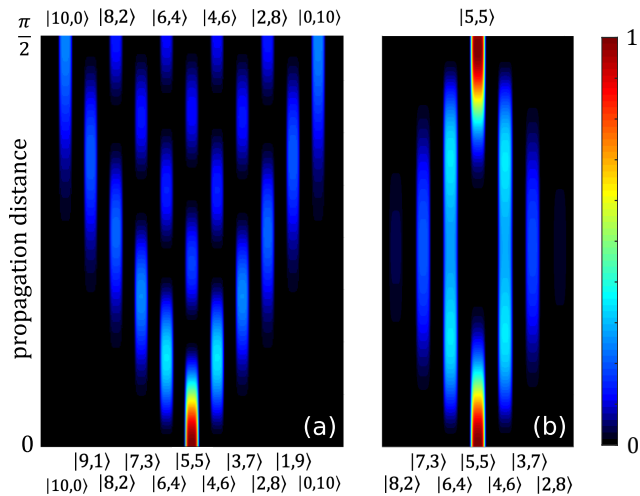}
   \caption{Probability distribution $\left|\bra{m,N-m} \hat{U}(z) \ket{\psi(0)}\right|^2$ for the initial state $\ket{\psi(0)}=\ket{5,5}$ propagating 
	          through a waveguide beam splitter with (a) $\beta_{1}=\beta_{2}=1$
						(discrete "diffraction" in state space) and (b) $\beta_{1}=0$ and 
						$\beta_{2}=4$ ("Bloch oscillations" in state space).}
\label{fig:Bloch}
\end{figure}
Computing the matrix elements $\hat{H}_{(m,n),(p,q)}=\bra{m,n}\hat{H}\ket{p,q}$ reveals 
that the states obey the $(N+1)$ equations of motion
\begin{align}
   \begin{split}
      i\frac{d\ket{m,n}}{dz} &= \left(\beta_{1}m+\beta_{2}n\right)\ket{m,n} \\
                             &+ C_{m}\ket{m-1,n+1} + C_{m+1}\ket{m+1,n-1},
   \end{split}
\label{eq:2M}
\end{align}
with $C_{m}=\kappa\sqrt{m(n+1)}$ and $n=N-m$ \cite{Tschernig:18}. This indicates that 
inside a waveguide beam splitter the amplitudes of two-mode $N$-photon states evolve 
coupled to each other with hopping rates $C_{m}$, and the corresponding phases depend 
on both propagation constants.\\
For the case of two identical waveguides we have $\beta_{1} = \beta_{2} = \beta$ so
that the first term on the r.h.s. of \eqref{eq:2M} becomes $\beta N\ket{m,N-m}$ 
which indicates that all the states will exhibit the same \emph{effective propagation 
constant}. 
Interestingly, it has been recently shown that waveguide beam splitters produce the 
Discrete Fractional Fourier Transform (DFrFT) of $N$-photon states 
\cite{Tschernig:18}, as well as exceptional points of arbitrary order, provided that losses are introduced 
in one of the waveguides \cite{Quiroz-Juarez:19}.\\ 
On the other hand, when considering two non-identical waveguides, $\beta_{1} \neq \beta_{2}$, the first term on the r.h.s. of \eqref{eq:2M} acquires the form 
$\left[(\beta_{1}-\beta_{2})m+\beta_{2}N\right]\ket{m,N-m}$. 
Remarkably, the term $\left[(\beta_{1}-\beta_{2})m\right]$ indicates that the state evolution will be influenced by an effective \emph{ramping potential} in the same 
fashion as in the case of classical waves in Bloch oscillator systems \cite{Silberberg1999,Keil2012,Lebugle2015}. 
Consequently, we can tailor the dynamics of $N$-photon states by simply adjusting 
the \emph{Bloch slope} $(\beta_{1}-\beta_{2})$ in order to suppress and/or create 
certain output states. As an illustration, we depict in \figref{fig:Bloch} the 
probability evolution for the initial state $\ket{5,5}$ in a waveguide beam splitter 
with coupling coefficient $\kappa=1$ (a) for $\beta_{1}=\beta_{2}=1$ and (b) for 
$\beta_{1}=0,\beta_{2}=4$. While case (a) corresponds to discrete "diffraction"
of the initial state in state space, case (b) corresponds to "Bloch oscillations"
in state space.
Note, that throughout this work we present all simulations using the normalized 
propagation coordinate $z=\kappa Z$, where $Z$ is the actual propagation distance 
and $\kappa$ stands for the nearest-neighbor coupling coefficient.
After the above introductory examples, we now proceed to consider the most interesting 
case where multiple photons $N>1$ excite more than two waveguides $M>2$. In order to 
motivate the concept of pseudo-energy we first examine the simplest case of a waveguide 
trimer, $M=3$, that is excited by $N=2$ photons and then move on to the general case. \\
For a waveguide trimer and two identical photons, the Hamiltonian takes the form
\begin{equation}
\begin{split}
   \hat{H} =&   \beta_1 \hat{a}^\dagger_1 \hat{a}_1 
	            + \beta_2 \hat{a}^\dagger_2 \hat{a}_2 
							+ \beta_3 \hat{a}^\dagger_3 \hat{a}_3 \\
            & + \kappa_1 \left(\hat{a}^\dagger_1 \hat{a}_2+\hat{a}^\dagger_2 \hat{a}_1 \right) 
						  + \kappa_2 \left(\hat{a}^\dagger_2 \hat{a}_3+\hat{a}^\dagger_3 \hat{a}_2 \right).
\end{split}
\end{equation}
In this scenario, we have a total of 6 photon-number states obeying the following 
coupled set of equations of motion
\begingroup
\allowdisplaybreaks
\begin{align}
   i \frac{d}{dz}\ket{200} &= 2\beta_1 \ket{200}+\sqrt{2}\kappa_1\ket{110}\label{eq:N2M3-first}\\
   i \frac{d}{dz}\ket{110} &= (\beta_1+\beta_2)\ket{110}+\kappa_2\ket{101}\\
                           &+ \sqrt{2}\kappa_1 \big(\ket{200}+\ket{020}\big)\\
   i \frac{d}{dz}\ket{020} &= 2\beta_2\ket{020}+\sqrt{2}\kappa_1\ket{110}+\sqrt{2}\kappa_2 \ket{011}\\
   i \frac{d}{dz}\ket{101} &= (\beta_1+\beta_3)\ket{101}+\kappa_1\ket{011}+\kappa_2\ket{110}\\
   i \frac{d}{dz}\ket{011} &= (\beta_2+\beta_3)\ket{011}+\kappa_1\ket{101} \\
                           &+\sqrt{2}\kappa_2 \big(\ket{002}+ \ket{020}\big)\\
   i \frac{d}{dz}\ket{002} &= 2\beta_3 \ket{002} + \sqrt{2}\kappa_2 \ket{011}.
\label{eq:N2M3-last}
\end{align}
\endgroup
As in the earlier examples, here we also have the possibility of molding 
the state dynamics via tuning the propagation constants and coupling coefficients. 
\begin{figure}[b!]
\centering
   \includegraphics[scale=.35]{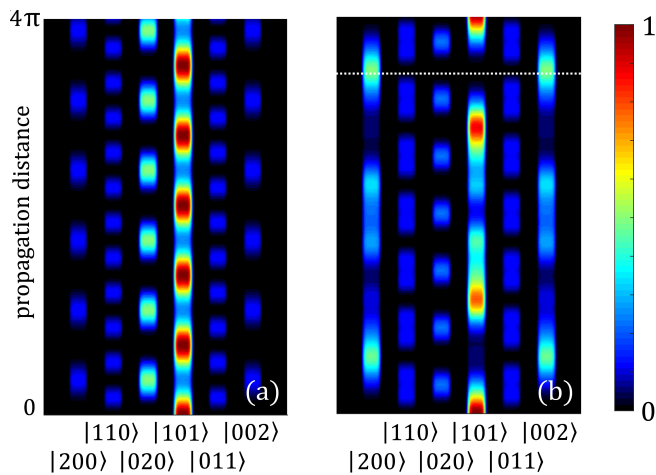}
   \caption{Probability distribution $\left|\bra{n_1,n_2,n_3} \hat{U}(z)\ket{\psi(0)} \right|^2$ for the initial state $\ket{\psi(0)}=\ket{1,0,1}$ propagating 
	          through a balanced 3-waveguide beam splitter ($\kappa_1=\kappa_2=1$) 
						with (a) $\beta_{1}=\beta_{2}=\beta_3=0$
						and (b) $\beta_{1}=\beta_3=0$ and $\beta_{2}=2$. 
						At the dotted horizontal line the state has evolved almost exactly 
						into a two-photon N00N state in state space.}
\label{fig:bloch101}
\end{figure}
For instance, for equal coupling coefficients $\kappa_1=\kappa_2=1$ and identical 
waveguides $\beta_1=\beta_2=\beta_3=0$, we observe a periodic spreading and contraction 
of the two-photon wave function, as illustrated in \figref{fig:bloch101}~(a). 
In contrast, choosing a different propagation constant for the central waveguide, 
$\beta_2=2$, leads to a quasi-periodic evolution, \figref{fig:bloch101}~(b).  
Indeed, this quasi-periodic evolution occurs because the ratios between the 
eigenvalues of the coupling matrix are irrational numbers. We would like to 
emphasize that at the propagation distance indicated by the dashed line in  
\figref{fig:bloch101}~(b), the input state $\ket{101}$ evolves into a quasi 
two-photon N00N state in state space, which is reminiscent of the Hong-Ou-Mandel 
effect 
\cite{Hong1987}.\\
To describe the photon dynamics in the waveguide trimer, we have obtained an 
even number of equations. At this point, the way in which the states should 
be arranged into a synthetic lattice is not at all clear. 
To be precise, the six states representing the sites of the synthetic lattice 
can be sorted into at least two distinct natural sequences as shown in 
Table~\ref{table:T1}.\\
\begin{table}[t!]
\begin{center}
   \begin{tabular}{|c|c|c|c|c|c|}
   \hline
   $\ket{2,0,0}$ $\ket{1,1,0}$ $\ket{0,2,0}$ $\ket{1,0,1}$ $\ket{0,1,1}$ $\ket{0,0,2}$\\
   \hline
   $\ket{2,0,0}$ $\ket{1,1,0}$ $\ket{1,0,1}$ $\ket{0,2,0}$ $\ket{0,1,1}$ $\ket{0,0,2}$\\
   \hline
\end{tabular}
\end{center}
  \caption{Possible lattice configurations for states arising in a waveguide 
	         trimer exited by two photons.}
\label{table:T1}
\end{table}
Clearly, arranging the states into a lattice (i.e., sorting) and analyzing 
the corresponding equations of motion becomes rather cumbersome when considering 
higher photon numbers in multiple coupled waveguides. In the following section, 
we, therefore, introduce a concise and universal method that facilitates studying 
the general case of $N>1$ photons propagating in arrays formed by $M>2$ waveguides. 
The resulting structures follow from physical and mathematical considerations 
that eventually allow us to describe multi-photon processes in waveguide arrays 
in a surprising and remarkable way that resembles the quantum-mechanical description 
of multi-level atoms.

\section{Pseudo energy representation}
\begin{figure}
\centering
   \includegraphics[scale=.3]{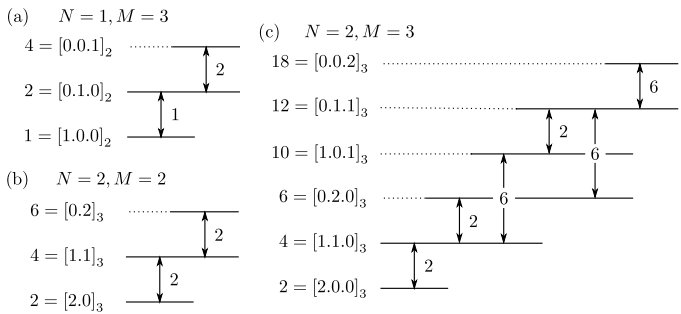}
   \caption{Pseudo-energy term diagrams for (a) $N=1$ photon in $M=3$ coupled 
	          waveguides, (b) $N=2$ photons in $M=2$ coupled waveguides, and 
						(c) $N=2$ photons in $M=3$ waveguides. Horizontal lines symbolize 
						the different Fock states and vertical arrows indicate allowed 
						transitions along with the corresponding pseudo-exchange energy. }
\label{fig:termdiagrams}
\end{figure}

\begin{figure}
\centering
   \includegraphics[scale=.6]{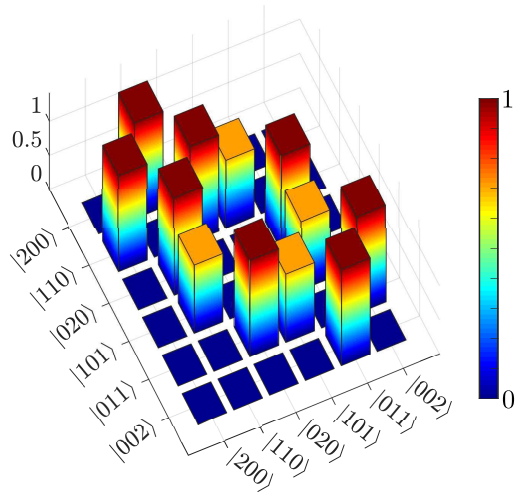}
   \caption{Matrix components of the effective Hamiltonian $H_{\mu\nu}$ for 
	          $N=2$ photons propagating in $M=3$ identical, nearest-neighbor 
						coupled waveguides ($\beta_1 = \beta_2$ and $\kappa_1=\kappa_2=1$).}
\label{fig:Hmatrix}
\end{figure}

We now introduce a concept analogous to the concept of energy and we, therefore,
refer to it as the \emph{pseudo-energy}. As we show below, the concept of 
\emph{pseudo-energy} is rather useful since it facilitates a unique sorting 
of multi-photon Fock states in a physically meaningful way and allows for
establishing a correspondence between Fock states and the energy levels of a 
\emph{synthetic atom}. 
Concurrently, we identify \emph{pseudo-energy ladder operators} along with
\emph{pseudo-exchange-energies} in order to define the corresponding \emph{selection 
rules} in Fock space for transitions between the \emph{pseudo-energy} levels
of the \emph{synthetic atom}. \\
We consider $N$ indistinguishable photons propagating in an array of $M$ 
lossless evanescently coupled waveguides that give rise to $N_F = (N+M-1)!/N!(M-1)!$ 
Fock states $\ket{n_1,\ldots,n_M}$, fulfilling the condition $\sum_{m=1}^{M}n_{m}~=~N$. 
The first issue to be addressed is to determine a way of sorting the multi-photon 
states in Fock space in a meaningful way. To do so, we associate a unique 
numerical value to every state $\ket{n_1,\ldots,n_M}$ as follows
\begin{align}
   \ket{n_1,\ldots,n_M} & \Rightarrow \left[n_1._{\ldots}.n_M \right]_{N+1}\nonumber\\
                        &=n_{1} \times (N+1)^0+_{\ldots}+n_{M }\times(N+1)^{M-1}.
\end{align}
Here, the subscript $N+1$ indicates that the numbers in the square brackets 
have to be expressed in base $N+1$, and the least significant digit is the 
left-most number $n_1$.
Observing that $\left[n_1._{\ldots}.n_M \right]_{N+1}=\sum_{m=1}^M \left(N+1 \right)^{m-1} n_m$ 
allows us to define the \emph{pseudo-energy operator}
\begin{equation}
   \hat{K}^{(N,M)}= \sum_{m=1}^M \left(N+1 \right)^{m-1} \hat{n}_m,
\end{equation}
such that its action on the $N$-photon-$M$-mode Fock states $\ket{n_1,\ldots,n_M}$ 
yields
\begin{equation}
\label{eq:Eig}
\hat{K}^{(N,M)} \ket{n_1,\ldots,n_M} =  K(n_1,\ldots,n_M)\ket{n_1,\ldots,n_M},
\end{equation}
with eigenspectrum $K(n_1,\ldots,n_M)=\sum_{m=1}^M \left(N+1 \right)^{m-1} n_m$. 
From \eqref{eq:Eig}, we readily infer the smallest and largest eigenvalues 
$K_{\text{min}} = K(N,0,\ldots,0,0) = \left[N.0._{\ldots}.0.0 \right]_{N+1} = N$ 
and 
$K_{\text{max}}=K(0,0,\ldots,0,N)=\left[0.0._{\ldots}.0.N \right]_{N+1}=N(N+1)^{M-1}$, 
respectively. 
Accordingly, the eigenvalues are bounded by $K_{\text{min}} \leq K_{\nu} \leq K_{\text{max}}$. \\
As a result, in order to sort the associated Fock states, we have to compute 
the corresponding $K_\nu$'s and arrange them in ascending order. The resulting 
ladder of $K_\nu$'s then defines the synthetic lattices formed by the states. 
We refer to this ordering as the \emph{pseudo-energy representation} of the 
$N$-photon-$M$-mode Fock states.\\ 
For illustration, we revisit the above case of $N=2$ photons propagating in an 
array of $M=3$ waveguides. Accordingly, there are $N_F = 6$ states and the 
spectrum of the pseudo-energy operator $\hat{K}^{(2,3)}$ comprises $6$ integers
\begin{align}
\label{eq:Eig2}
   & \left\{ \left[2.0.0 \right]_{3},
	           \left[1.1.0 \right]_{3},
			       \left[0.2.0 \right]_{3},
				  	 \left[1.0.1 \right]_{3}, 
						 \left[0.1.1 \right]_{3},
						 \left[0.0.2 \right]_{3}
		 \right\}\nonumber\\
   & = \{2,4,6,10,12,18\}.
\end{align}
Using these numbers we readily obtain the pseudo-energy representation of the 
2-photon-3-mode Fock space
\begin{align}
\begin{split}
   \ket{2,0,0}  &= \ket{\left[2.0.0 \right]_{3}=2}  = \ket{K_1},\\
   \ket{1,1,0}  &= \ket{\left[1.1.0 \right]_{3}=4}  = \ket{K_2},\\
   \ket{0,2,0}  &= \ket{\left[0.2.0 \right]_{3}=6}  = \ket{K_3},\\
   \ket{1,0,1}  &= \ket{\left[1.0.1 \right]_{3}=10} = \ket{K_4},\\
   \ket{0,1,1}  &= \ket{\left[0.1.1 \right]_{3}=12} = \ket{K_5},\\
   \ket{0,0,2}  &= \ket{\left[0.0.2 \right]_{3}=18} = \ket{K_6}.
\end{split}
\end{align}
Consequently, we designate $K_{\nu}$ as the \emph{pseudo-energy} of the 
$\nu$-th Fock state in the $N$-photon-$M$-mode Fock space
\begin{equation}
   \left| K_{\nu} \right\rangle = \left| \left[ n_1^{(\nu)}._{\ldots}.n_M^{(\nu)} \right]_{N+1}\right\rangle 
	                              = \left|n^{(\nu)}_1,\ldots,n^{(\nu)}_M \right\rangle,
\end{equation}
with $\nu=1,\ldots,N_F$. 
In general, for any given $N,M$ and pseudo-energy $K_{\nu}$, the inverse 
mapping onto the mode-occupation numbers is
\begin{equation}
   n^{(\nu)}_m = \left(K_{\nu} \div (N+1)^{m-1}\right) \# (N+1),
\end{equation}
where the symbol $\div$ corresponds to integer division and $\#$ is 
the modulo operator. \\
We now proceed to show how the pseudo energy representation of Fock states 
allows us to express the equations of motion of $N$ photons in $M$ waveguides 
in a concise way. To do so, we take a closer look at the action of the 
operator $\hat{a}^\dagger_i \hat{a}_j$ on a Fock state
\begin{equation}
\label{eq:actionij}
\begin{split}
   \hat{a}^\dagger_i \hat{a}_{j} \ket{n_1,\ldots,n_M} = \sqrt{(n_i+1)n_j} &\Big|n_1,\ldots,n_i+1,\\
                                                                          &\ldots,n_j-1,\ldots,n_M \Big\rangle.
\end{split}
\end{equation}
If the state $\ket{n_1,\ldots,n_M}$ corresponds to the pseudo-energy $K_\nu$, 
then the resulting state on the r.h.s. of \eqref{eq:actionij} must have 
the pseudo-energy
\begin{align}
\begin{split}
   K_\mu & = \left[n_1._{\ldots}.n_i+1._{\ldots}.n_j-1._{\ldots}.n_M \right]_{N+1} \\
         & = K_\nu + (N+1)^{i-1} - (N+1)^{j-1}.
\end{split}
\end{align}
Therefore, the action of $\hat{a}^\dagger_i \hat{a}_j$ changes the pseudo-energy 
of Fock states by the amount
\begin{equation}
\Delta K_{ij} = (N+1)^{i-1} - (N+1)^{j-1}=-\Delta K_{ji},
\end{equation}
which we denote as the \emph{pseudo-exchange energy} associated with the tunneling 
process taking place between waveguides $i$ and $j$. In this sense the operators 
$\hat{a}^\dagger_i \hat{a}_j$ can be thought of as \emph{pseudo-energy ladder 
operators}, which raise or lower the pseudo-energy of Fock states. Consequently, 
we can write
\begin{align}
   \bra{K_\mu}\kappa_{ij}\hat{a}^\dagger_i \hat{a}_j\ket{K_\nu} 
	 & =
	 \kappa_{ij} \sqrt{\left(n^{(\nu)}_i+1\right)n^{(\nu)}_j} \delta_{K_\mu,K_\nu+\Delta K_{ij}}.
\label{eq:effectivecoupling}
\end{align}
The physical significance of \eqref{eq:effectivecoupling} is that a direct 
transition between the states $\ket{K_\mu}$ and $\ket{K_\nu}$ is only possible 
if there exists a pseudo-exchange energy $\Delta K_{ij}$ such that 
\begin{equation}
\label{eq:selectionrule}
   |\Delta K_{ij}|=|K_\mu -K_\nu|.
\end{equation}
Obviously, \eqref{eq:selectionrule} defines the \emph{selection rules} in 
Fock space. Together with the action of the photon number operators $\hat{n}_m$, 
the full system of coupled equations governing the propagation of $N$ photons 
through $M$ coupled waveguides in the pseudo-energy representation is given by
\begin{equation}
\begin{split}
  i\frac{d}{dz} \ket{K_{\mu}} &= \sum_{m=1}^M \beta_m n_m^{(\mu)} \ket{K_{\mu}} \\
                              &+ \sum_{\nu=1}^{N_F} 
															   \sum_{i,j=1}^M 
																 \kappa_{ij} 
																 \sqrt{\left(n^{(\nu)}_i+1\right)n^{(\nu)}_j} 
																 \delta_{K_\mu,K_\nu+\Delta K_{ij}} 
																 \ket{K_\nu}.
\end{split}
\end{equation}
For the case of nearest-neighbour coupled, identical waveguides, where all the 
propagation constants are the same, the relevant pseudo-exchange energies are 
$\Delta K_i = \Delta K_{i+1,i}=N(N+1)^{i-1}$ and the set of coupled equations 
reduces to
\begin{align}
\label{eq:matrix-representation}
\begin{split}
   i\frac{d}{dz} \ket{K_{\mu}} & = N \beta \ket{K_\mu} \\
                               & + \sum_{\nu=1}^{N_F}  \sum_{i=1}^{M-1}
															     \kappa_i \Bigg(\sqrt{\left(n^{(\nu)}_i+1\right)n^{(\nu)}_{i+1}} 
																	 \delta_{K_\mu,K_\nu-\Delta K_i} \\
                               & + \sqrt{n^{(\nu)}_i\left(n^{(\nu)}_{i+1}+1\right)} 
															     \delta_{K_\mu,K_\nu+\Delta K_i} \Bigg) 
																	 \ket{K_\nu}.
\end{split}
\end{align}
To further illustrate the resulting coupling system in Fock space, we revisit the 
case of a single photon $N=1$ propagating in $M=3$ waveguides. The effective coupling 
behavior - of allowed and forbidden transitions in Fock space - can now be visualized 
within a \emph{pseudo-energy term diagram}, as illustrated in \figref{fig:termdiagrams} (a). 
In this particular case the nearest-neighbour coupling of the waveguides is retained 
in Fock space and any given Fock state $\ket{K_\nu}$ only couples to its nearest 
neighbours $\ketr{K_{\nu\pm 1}}$. \\
A similar picture arises in the case of two waveguides $M=2$ and $N=2$ photons, as 
depicted in \figref{fig:termdiagrams} (b). Here, we obtain a term-diagram that is 
essentially isomorphic to \figref{fig:termdiagrams} (a), where -- again -- only 
nearest-neighbour Fock states are coupled to each other.\\
The nearest neighbour picture radically changes when applying the pseudo-energy 
approach to the case of $N~=~2$ photons and $M=3$ waveguides as displayed in the 
corresponding term-diagram in \figref{fig:termdiagrams} (c). Importantly, even 
when the waveguides are -- in real space -- only coupled to their nearest 
neighbors, in photon number space certain states become coupled to next-nearest 
neighbor states. 
For instance, in \figref{fig:termdiagrams} (c) we observe that the state 
$\ketr{K_2}=\ketr{4}=\ket{1,1,0}$ not only couples to its neighbors 
$\ketr{K_1}=\ketr{2}=\ket{2,0,0}$ and $\ketr{K_3}=\ketr{6}=\ket{0,2,0}$, but 
also to the next-nearest neighbour state $\ketr{K_4}=\ketr{10}=\ket{1,0,1}$. 
For illustrative purposes, we present in \figref{fig:Hmatrix} the coupling 
matrix for this particular set of states when the three-waveguide system is 
formed by identical waveguides, $\beta_1=\beta_2=\beta_3=0$, and balanced coupling 
coefficients $\kappa_1=\kappa_2=1$.\\
At this point is rather evident that the richness and complexity of the emerging 
synthetic configurations will become more prominent when higher number of photons 
and waveguides are considered. Moreover, it is worth stressing that in order to 
generate the present synthetic structures we did not require any modulation of 
the system parameters as the states naturally couple due to the system's internal
dynamics.\\
\section{Non-planar synthetic lattices: Fock graphs}
\begin{figure*}
\centering
   \includegraphics[scale=.35]{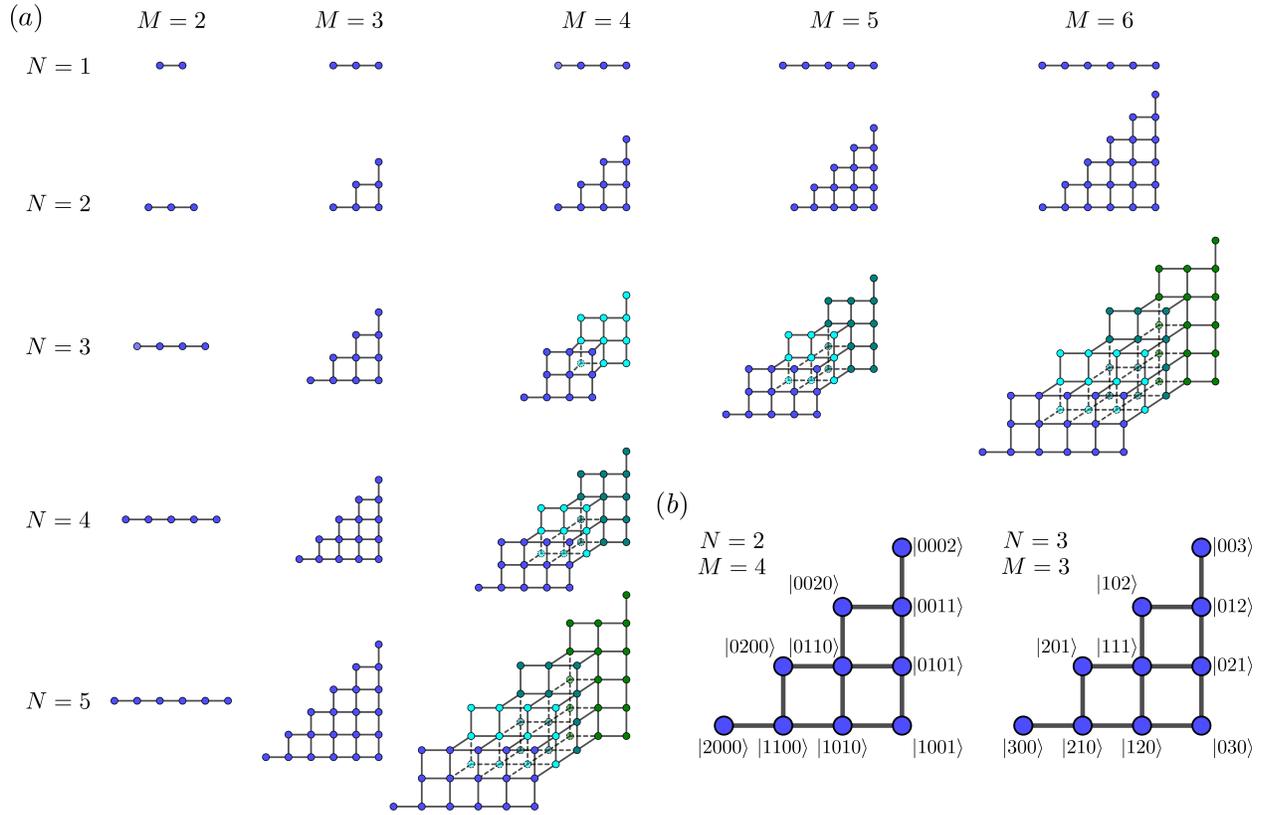}
   \caption{(a) Overview of several two- and three-dimensional embeddings of Fock graphs $A^{(N,M)}_{\mu,\nu}$ for $M=2,\ldots,6$ waveguides excited by $N=1,\ldots,5$ indistinguishable photons. Different node colors indicate layer-like structures that emerge for $N\geq3,M\geq4$ (all nodes in the same layer feature the same color). For the sake of readibility we have omitted the node labels as well as the graphs for $M\geq 5,N\geq 4$. (b) The smallest example of an isomorphic pair of planar Fock graphs with $N=2,M=4$ and $N=3,M=3$ respectively.}
\label{fig:fockgraphs}
\end{figure*}

In this section we introduce a more convenient way of representing the hamiltonian 
matrix of $N$-photons exciting $M$-waveguides. To do so, we interpret the states as 
\emph{vertices} of a \emph{graph} (\emph{Fock graph}) where the allowed inter-state 
transitions represent the \emph{edges}. 
A practical representation of finite graphs is the so-called adjacency matrix whose 
entries indicate whether pairs of vertices are adjacent or not. In the present context, 
the effective Hamiltonian $H_{\mu\nu}=\bra{K_\mu} \hat{H} \ket{K_\nu}$ in the 
$N$-photon-$M$-mode pseudo-energy representation determines such an adjacency matrix 
\begin{equation}
   A^{(N,M)}_{\mu\nu}= \Theta(H_{\mu\nu}),
\label{eq:adjacencymatrix}
\end{equation}
where $\Theta$ is the step-function and $A^{(N,M)}_{\mu\nu} = 1 (\mbox{or } 0)$ indicates 
a connection (or no connection) between the vertices $\mu$ and $\nu$. In what follows, 
we assume identical waveguides with $\beta_1 =...= \beta_M=0$ in order to omit self-loops 
in the graph representation. As an example, in \figref{fig:N2M3graph} we depict the Fock 
graph arising from the effective Hamiltonian of \figref{fig:Hmatrix}, which we have 
already discussed in the previous section. 
In \figref{fig:fockgraphs}~(a), we depict further examples for photon numbers up to $N=5$ 
and up to $M = 6$ waveguides. The first row, which corresponds to single photon graphs, 
simply reflects the one-dimensional spatial configuration of the waveguides. By introducing 
a second photon, we observe that the Fock graphs become two-dimensional, \figref{fig:fockgraphs}~(b), 
except for the case $M=2$. The inclusion of more photons leads to non-planar graphs, i.e. graphs that cannot be drawn in 2D without intersecting edges, which 
exhibit a layered structure in three dimensions as indicated by the different coloring 
of the nodes in different layers.\\
A prominent feature to highlight is the symmetry observed among graphs emerging for 
the combinations $(M,N)$ and $(M-l,N+l)$ and for $(M,N)$ and $(M+l,N-l)$, where $l$ 
is an integer.
%This symmetry is observed except when $M=N+1$ .
%A prominent feature is the symmetry with respect to the $M=N+1$ diagonal. 
In other words, every Fock graph has an isomorphic partner graph
\begin{equation}
   A_{\mu\nu}^{(N,M)} = A_{\mu\nu}^{(M-1,N+1)} \ \forall \ N,M,
\label{eq:graphpairs}
\end{equation}
with an identical adjacency matrix, up to a trivial permutation of the node labels. 
%We provide no proof for this statement, but we were able to verify this property numerically for all Fock graphs with $N \leq M \leq 10$. 
%As a consequence, the Fock graphs $A_{\mu\nu}^{(N,N+1)}$ are - in this sense - unique. 
In \figref{fig:fockgraphs} (b), we depict the smallest non-trivial pair of Fock 
graphs and the corresponding adjacency matrices that are induced by the pseudo-energy 
representation. If we were to start from $A^{(3,3)}_{\mu\nu}$ and permute its rows 
and columns according to $\left(1,\ldots,10 \right) \rightarrow \left(1,2,4,7,3,5,8,6,9,10\right)$ 
we will exactly obtain $A^{(2,4)}_{\mu\nu}$. \\
Indeed, this underlying symmetry in the space of possible Fock graphs has very 
interesting implications. For instance, in 
Ref. \cite{Tschernig:18} 
we have shown that it is possible to implement the number-resolved $N+1$-dimensional 
Discrete Fractional Fourier Transform (DFrFT) with a single waveguide beam splitter by 
launching $N$ indistinguishable photons. Furthermore, using the same photon-number-resolved 
mapping in Ref.~\cite{Quiroz-Juarez:19} 
we have shown how to attain so-called exceptional points of $N+1$ order, by way 
of exciting a semi-lossy waveguide beam splitter with high photon number states. 
In fact, it is now clear that these results emerge as special cases of 
\eqref{eq:graphpairs}, which pertains to the identity of the first row and 
column in \figref{fig:fockgraphs} (a). Thus, by following similar ideas it is 
possible, in principle, to find the corresponding effects for waveguide systems 
with $M\geq 3$ excited by $N\geq 2$ photons. \\
Additionally, by exploiting the graph symmetry it becomes apparent that a specific 
transformation which requires $N$ photons and $M$ waveguides could likewise be
implemented with $M-1$ photons and $N+1$ waveguides. Of course, such alternative
pathways of implementing a transformation are not always guaranteed because of 
the different dimensions of the experimentally accessible parameter spaces. 
Nonetheless, this may serve as a useful Ansatz to overcome concrete experimental 
difficulties. \\
Quite interestingly, synthetic Fock lattices have been  
explored previously in the context of QED circuits by Wang et al. \cite{Wang2016}.
In such a study, the joint excitation states of an atom coupled to the $N$-photon $3$-cavity 
Fock space form a two-dimensional, hexagonal Haldane-like synthetic lattice, which facilitates 
the generation of high photon-number NOON states. Crucially, the realization of this scheme demands the judicious implementation of the coupling between atom and cavity, as well as the precise modulation of the cavity resonance frequencies.
In contrast, the multi-photon synthetic dimensions explored in the present work are intrinsically active by virtue of the indistinguishability of the photons,
and as such they do not require any external driving of the system's parameters. \\
The Fock graphs offer a rich variety of synthetic coupled structures. This variety 
can be further enhanced by considering different spatial arrangements of the 
waveguides, for instance, ring- or star-shaped structures instead of the simple 
planar configuration studied here. 
Importantly, the evolution of multi-photon states in synthetic lattices and graphs can be dynamically reconfigured by using programmable photonic chips \cite{Shadbolt2012}. That is, integrated optical devices where the waveguides' refractive index and coupling coefficients can be modified externally.
Nevertheless, even with this simple one-dimensional arrangement comprising a few waveguides, small photon numbers, and a time-independent Hamiltonian, 
one encounters interesting effects that are only possible due to the multi-dimensionality of the corresponding Fock graphs. 
\section{All-optical dark states and parallel quantum random walks}
To show possible applications of the pseudo-energy synthetic lattices we discuss the generation of all-optical dark states \cite{Lambropoulos2007} and multi-photon quantum. The simplest dark states are encountered in 3-level atomic- or molecular systems, where radiative transitions between, e.g.,  
$\ket{1} \leftrightarrow \ket{2} \leftrightarrow \ket{3}$ are allowed but the transition
$\ket{1} \leftrightarrow \ket{3}$ is forbidden. In this simple scenario, a dark state is
a superposition of the uncoupled states $\ket{D}=\cos (\theta) \ket{1} - \sin (\theta) \ket{3}$, 
where $\theta$ is given in terms of the Rabi frequencies of the allowed transitions \cite{Lambropoulos2007}. 
Once the system is in such a state, adiabatic changes in the Rabi frequencies allow for the tuning of 
the populations of the states $\ket{1}$ and $\ket{3}$, while the probability of $\ket{2}$ remains 0.
This ineresting behavior, which seemingly evades the radiative selection rules, can be mimicked in the pseudo-energy representation of Fock states using our all-optical setup. \\
\begin{figure}
\centering
   \includegraphics[scale=1.3]{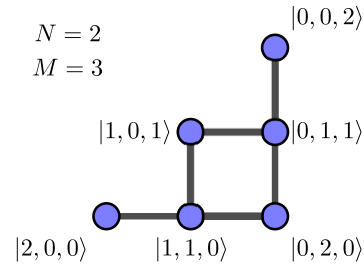}
   \caption{Two-dimensional Fock graph for $M=3$ waveguides excited by $N=2$ indistinguishable photons. The corresponding adjacency matrix is induced by the effective Hamiltonian in \figref{fig:Hmatrix} according to \eqref{eq:adjacencymatrix}.}
\label{fig:N2M3graph}
\end{figure}
To do so, we revisit one more time the case of $M=3$ waveguides, 
with equal propagation constants $\beta_1=\beta_2=\beta_3=0$ and balanced 
coupling coefficients $\kappa_1=\kappa_2=\frac{1}{\sqrt{2}}$, excited by 
$N=2$ photons. The pseudo-energy representation of the effective Hamiltonian 
takes the form
\begin{equation}
   H_{\mu\nu} = \begin{pmatrix}
                   0   &         1        &   0   &         0        &         0        &   0\\
                   1   &         0        &   1   &\frac{1}{\sqrt{2}}&         0        &   0\\
                   0   &         1        &   0   &         0        &         1        &   0\\
                   0   &\frac{1}{\sqrt{2}}&   0   &         0        &\frac{1}{\sqrt{2}}&   0\\
                   0   &         0        &   1   &\frac{1}{\sqrt{2}}&         0        &   1\\
                   0   &         0        &   0   &         0        &         1        &   0
                \end{pmatrix}.
\end{equation}
With this choice of parameters the spectrum of $H_{\mu\nu}$ is integer-valued
\begin{equation}
   \left(\lambda_1,\ldots,\lambda_6  \right)=\left( -2,-1,0,0,1,2 \right),
\end{equation}
which indicates that the third and fourth eigenstates are degenerate with 
eigenvalues $\lambda_3=\lambda_4=0$. We now consider the evolution of a 
coherent superposition $\ket{\psi}$ of the eigenstates
\begin{equation}
  \ket{\phi_3} = \begin{pmatrix}
                    \frac{1}{2}    \\
                         0         \\
                         0         \\
                -\frac{1}{\sqrt{2}}\\
                         0         \\
                   \frac{1}{2}
                 \end{pmatrix} 
								\text{ and } 
 \ket{\phi_5} = \frac{1}{2} 
                \begin{pmatrix}
                        1          \\
                        1          \\
                        0          \\
                        0          \\
                       -1          \\
                       -1
                \end{pmatrix}
\end{equation}
with corresponding eigenvalues $\lambda_3=0$ and $\lambda_5=1$, specifically
\begin{equation}
   \ket{\psi} = \frac{1}{\sqrt{2}} \left(\ket{\phi_3}+\ket{\phi_5} \right) 
	            = \frac{1}{\sqrt{2}}
							  \begin{pmatrix}
                       1           \\
                \frac{1}{2}        \\
                       0           \\
             -\frac{1}{\sqrt{2}}   \\
               -\frac{1}{2}        \\
                       0
               \end{pmatrix}.
\end{equation}
In the standard Fock representation $\ket{\psi}$ reads as
\begin{equation}
   \ket{\psi} = \frac{1}{\sqrt{2}}\left(\ket{200}+\frac{1}{2}\ket{110}-\frac{1}{\sqrt{2}}\ket{101}-\frac{1}{2}\ket{011} \right).
\label{eq:superposition}
\end{equation}
\begin{figure}
\centering
   \includegraphics[scale=.4]{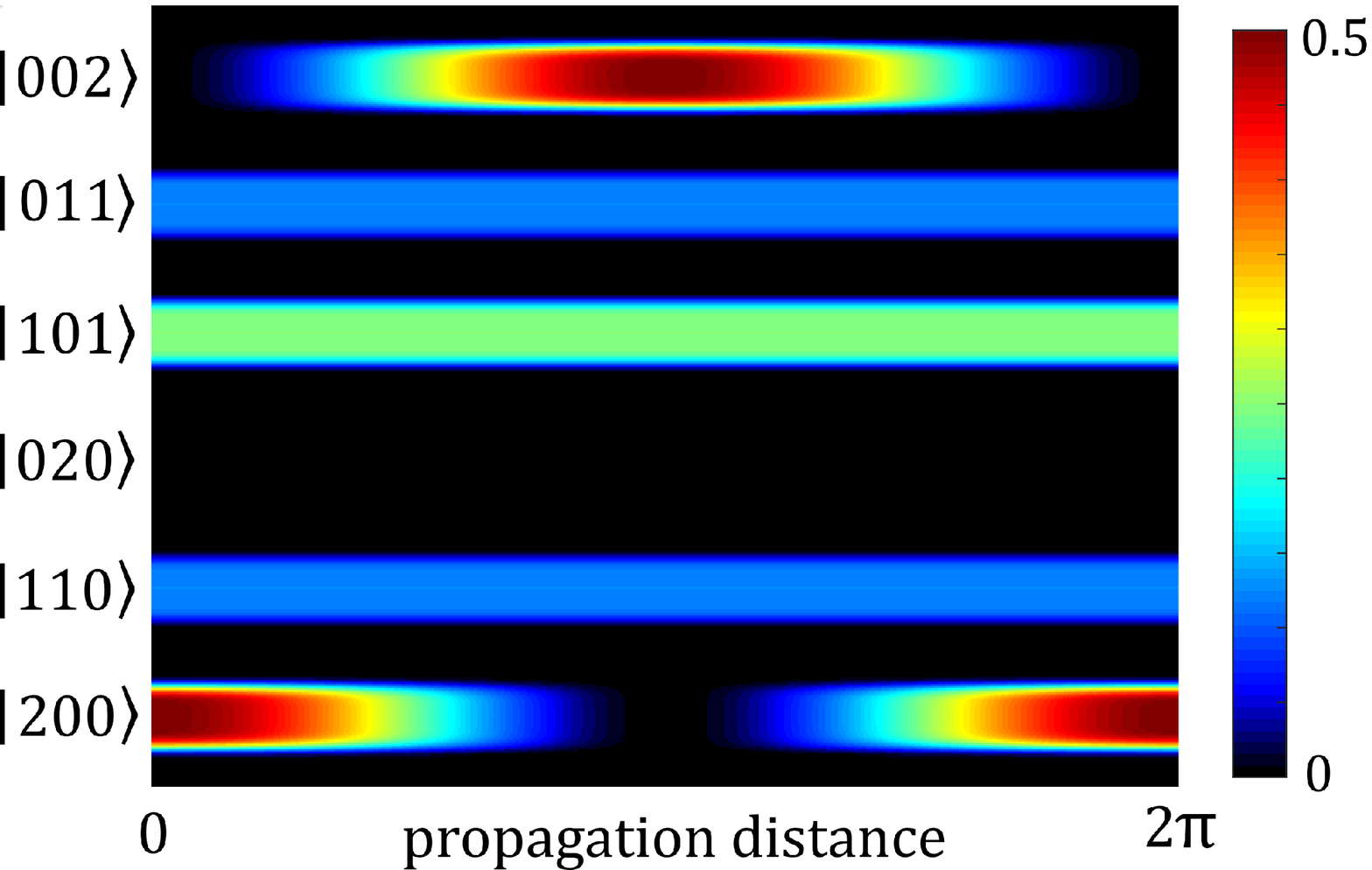}
   \caption{Evolution of the probabilities $\left|\bra{K_\nu} \hat{U}(z) \ket{\psi} \right|^2$ of the state $\ket{\psi}$ 
	          as defined in \eqref{eq:superposition}.}
\label{fig:qpsN2M3}
\end{figure}
The probability evolution for this state is shown in \figref{fig:qpsN2M3}. 
As one can see, this state displays the characteristic behavior of a dark state. That is, the initial state evolves exhibiting oscillating transitions between the states $\ket{200}$ and $\ket{002}$ with period $\frac{2\pi}{\lambda_5-\lambda_3} = 2\pi$. These transitions occur in spite of the fact that direct transition $\ket{200}\to \ket{002}$ is forbidden $\left(\bra{200}\hat{H}\ket{002}=0\right)$, and  those states have the maximum 
possible distance within the graph, that is, at least 4 single-photon tunneling processes are required to transform one state into the other. All probabilities of the intermediate states remain constant and, in a way, assist the simultaneous tunneling of two photons between the outermost waveguides. We stress that this 6-level dark-state state is induced by a time-independent hamiltonian and it occurs naturally without the need of adiabatic fine-tuning of the parameters. We would further like to note, that the 
state $\ket{020}$ exhibits zero probability for all $z$, further attesting a 
multi-photon tunneling (in this case co-tunneling) effect taking place between 
the two waveguides. 
Geometrically speaking, this effect arises due to destructive interference taking 
place in the two-way branching of the Fock graph shown in \figref{fig:N2M3graph}. 
This branching effectively allows for the flow of the amplitudes to take a 
`detour'\ around the $\ket{020}$ node. \\
As an alternative, one may attempt to implement a real space structure in one
or two dimensions consisting of six coupled waveguides in order to emulate an 
equivalent Hamiltonian for just a single photon. However, this would be topologically
impossible, since there always exist additional cross-talk between the waveguides 
representing the nodes at the center of the graph. In other words, our Fock-graph
based analysis of multi-photon propagation in waveguide arrays allows the realization
of functionalities beyond what can be realized with linear (single-photon) based
networks. \\
Quite interestingly, by exciting waveguide lattices with multi-photon states 
comprising infinite coherent superpositions, e. g. coherent states 
$\ket{\alpha}=\exp\left(-|\alpha|^{2}/2\right)\sum_{n=0}^{\infty}\left(\alpha^{n}/\sqrt{n!}\right)\ket{n}$ 
or two-mode squeezed vacuum states $\ket{\xi}=\sqrt{1-|\xi|^{2}}\sum_{n=0}^{\infty}\xi^{n}\ket{n,n}$, 
opens a route to generating, in principle, an infinite number of lattices or 
graphs with different numbers of lattice sites and many dimensions simultaneously. 
This possibility is very appealing for realizing parallel quantum random walks 
where the corresponding walkers can perform different numbers of steps that 
depend on the number of photons involved in each process. We stress that the 
observation of these quantum walks is nowadays possible utilizing bright 
parametric down-conversion sources in combination with photon-number-resolving 
detectors \cite{Magana2019}.

\section{Eigendecomposition in the pseudo-energy representation}
In this final section, we obtain an analytical expression for the eigensystem 
of an $M$ waveguide system (or tight-binding network) with arbitrary coupling 
coefficients $\kappa_{m}$ excited by $N$ indistinguishable photons. With the 
help of the pseudo-energy representation we will be able to find a concise 
expression, which also introduces a natural ordering of the $N$-photon-$M$-waveguide 
eigenstates. As we have seen, in the case of a single photon $N=1$, the Hamiltonian 
takes on a bi-diagonal form in the pseudo-energy representation. In some cases 
it is possible to find an analytical closed form expression for the eigensystem, 
as for example in the case of the DFrFT 
\cite{Weimann2016}. 
Even if no analytical solution is available, numerical algorithms are known 
\cite{Gill1990}, that deal with bi-diagonal matrices efficiently. Therefore, 
without loss of generality we assume that we know the complete eigensystem of 
the single-photon-$M$-waveguide Hamiltonian, which we denote as
\begin{align}
           \ket{\phi_n} &= \sum_{m=1}^{M} u^{(n)}_m \hat{a}^\dagger_{m} \ket{0} = \sum_{m=1}^{M} u^{(n)}_m \ketr{K_m},\\
   \hat{H} \ket{\phi_n} &= \lambda_n \ket{\phi_n},
\end{align}
where $n=1,\ldots,M$. In the above equation, $u^{(n)}_m$ is the $m$-th component 
of the $n$-th eigenvector of the matrix $\hat{H}_{m,n}=\brar{K_m} \hat{H} \ketr{K_n}$ 
and it defines the single-particle eigenstates
\begin{equation}
   \hat{\phi}^\dagger_n = \sum_{m=1}^M u^{(n)}_m \hat{a}^\dagger_{m}.
\end{equation}
When the same waveguide system is excited by $N~>~1$ photons, it is clear that 
the many-particle eigenstates arise from the tensor-products of the single-particle 
eigenstates. Formally, we may write the resulting states as
\begin{equation}
   \ket{\tilde{n}_1,\ldots,\tilde{n}_M} = \prod_{m=1}^M \hat{\phi}^{\dagger \tilde{n}_m}_m \ket{0},
\end{equation}
but now the occupation numbers $\tilde{n}_m$ pertain to the number of photons 
occupying the $m$-th single-particle eigenmode. Consequently, we can apply the 
pseudo-energy ordering to the $N$-particle eigenstates by defining 
$\tilde{K}_\nu = \left[\tilde{n}^{(\nu)}_1._{\ldots}.\tilde{n}^{(\nu)}_M \right]_{N+1}$. 
The $\nu$-th eigenstate of the $N$-photon system is then given by
\begin{equation}
\label{eq:Nphoton-eigenstates}
  \ketr{\tilde{K}_\nu} = \prod_{m=1}^M \left(\sum_{k=1}^M u^{(m)}_k \hat{a}^\dagger_k \right)^{\tilde{n}_m^{(\nu)}} \ket{0}.
\end{equation}
Note, that in most cases it is necessary to normalize the resulting expression on 
the r.h.s. of \eqref{eq:Nphoton-eigenstates}. 
By requiring $\ketr{\tilde{K}_\nu}=\sum_{\mu=1}^{N_F} c^{(\nu)}_\mu \ketr{K_\mu}$, 
where $\ketr{K_\mu}$ denotes $N$-photon-$M$-waveguide Fock states, we find for 
the components $c^{(\nu)}_\mu$
\begin{equation}
  c^{(\nu)}_{\mu}=\brar{K_\mu} \prod_{m=1}^M \left(\sum_{k=1}^M u^{(m)}_k \hat{a}^\dagger_k \right)^{\tilde{n}_m^{(\nu)}} \ket{0}.
\end{equation}
It is now rather straightforward to show, that the $N$-particle eigenvalues are 
given as the sum of the eigenvalues of the involved single-particle eigenstates
\begin{equation}
\label{eq:NphotonEigenvalues}
   \tilde{\lambda}_\nu = \sum_{m=1}^M \tilde{n}_m^{(\nu)} \lambda_m.
\end{equation}
Using \eqref{eq:Nphoton-eigenstates} and \eqref{eq:1-M} it is 
straightforward to find the $N$-photon-$M$-waveguide time-evolution operator 
$\hat{U}(t)=\sum_{\nu=1}^{N_F} e^{-i\tilde{\lambda}_\nu t} \ket{\tilde{K}_\nu}\bra{\tilde{K}_\nu}$. 
We would like emphasize that the numerical evaluation of \eqref{eq:Nphoton-eigenstates} 
is far more efficient than the direct diagonalization of the full matrix representation 
of $\hat{H}$ in $N$-photon-$M$-waveguide Fock space. Due to the size and highly 
non-trivial structure of the resulting matrices, general eigensystem-solvers produce 
a significant amount of overhead, which we avoid in our approach. Essentially, we 
do not even require a calculation of the full matrix representation $H_{\mu\nu}$.
Instead, knowledge of the single-particle eigensystem and the bosonic nature of 
photons suffices.

\section{Conclusion}
In summary, we have shown that the propagation of multi-photon states through 
multi-port waveguide systems (tight-binding networks) gives rise to multiple 
synthetic lattices and multi-dimensional Fock graphs that allow for transparent
analyses of the relevant physical processes and the design of novel functionalities
beyond the linear (single-photon) realm. Since such synthetic structures emerge 
in the photon-number space we have been able to associate coherent multi-photon 
processes to parallelized multi-dimensional quantum random walks. 
This parallelization brings about novel opportunities for the implementation
of random walks where the randomness is not only present in the dynamics of the 
walkers but also in the simultaneous occurrence of different walks.

\section*{Acknowledgments}
We acknowledge support by the Deutsche Forschungsgemeinschaft (DFG) within
the framework of the DFG priority program 1839 {\it Tailored Disorder}.

% Bibliography
\bibliographystyle{unsrt}
%\bibliography{literature}
\providecommand{\noopsort}[1]{}\providecommand{\singleletter}[1]{#1}%

\end{document}